\documentclass[amssymb,useAMS,twocolumn,aps,prd,longbibliography,nofootinbib,superscriptaddress,amsmath]{revtex4-1}
\usepackage{graphicx}
\usepackage{amssymb}
\usepackage{wasysym}
\usepackage[caption=false]{subfig}
\usepackage{array, xcolor}
{

\usepackage[colorlinks=true,linkcolor=blue,citecolor=blue,urlcolor=black]{hyperref}

\begin{document}
\title{Probing sub-GeV mass SIMP dark matter with a low-threshold surface experiment}
\author{Jonathan H. Davis}
\affiliation{Theoretical Particle Physics and Cosmology, Department of Physics, King's College London, London WC2R 2LS, United Kingdom
\\ {\smallskip \tt  \href{mailto:jonathan.davis@kcl.ac.uk}{jonathan.davis@kcl.ac.uk}}}
\date{\today}
\begin{abstract}
Using data from the $\nu$-cleus detector, based on the surface of the Earth, we place constraints on dark matter in the form of Strongly Interacting Massive Particles (SIMPs) which interact with nucleons via nuclear-scale cross sections.
For large SIMP-nucleon cross sections the sensitivity of traditional direct dark matter searches using underground experiments is limited by the energy loss experienced by SIMPs, due to scattering with the rock overburden and experimental shielding on their
way to the detector apparatus. Hence a surface-based experiment is ideal for a SIMP search, despite the much larger background, resulting from the lack of shielding. We show using data from a recent surface run of a low-threshold cryogenic detector that values of the SIMP-nucleon cross section up to approximately $10^{-27}$~cm$^2$ can be excluded for SIMPs with masses
above 100~MeV.
\end{abstract}

\maketitle

\section{Introduction}
There is strong evidence that the majority of matter in the Universe is in the form of so-called dark matter (DM)~\cite{Ade:2015xua}, whose presence is inferred via its gravitational interactions with luminous matter (which makes up stars and galaxies),
but which does not significantly scatter~\cite{Davis:2014zda} or emit radiation. Since the luminous matter in the Universe is composed of particles, specifically those of the Standard Model, it is reasonable to assume that the 
dark matter is also made of particles, albeit of a so-far undiscovered species.
Most searches for dark matter particles operate under the reasonable assumption that they interact only weakly with ordinary matter, with many searches focusing on so-called Weakly Interacting Massive Particles (WIMPs).
Hence, for example, direct dark matter search experiments are placed deep underground in order to vastly reduce
the background from the visible sector, such as cosmic rays, while leaving unaffected any potential signal from  WIMPs~\cite{Aprile:2017iyp,Akerib:2016vxi,Agnese:2015nto,Tan:2016zwf}. The majority of these searches have low-energy thresholds 
around a keV, and so are sensitive mostly to dark matter particles with masses above a GeV, though much recent progress has been made on lowering this threshold and probing lighter dark matter~\cite{Knapen:2016cue,Angloher:2017sxg,Strauss:2017cuu,Strauss:2017cam,Angloher:2015ewa,McCabe:2017rln,Kouvaris:2016afs,Agnese:2017jvy,Hochberg:2016sqx}.

However WIMPs are not the only potential dark matter candidate. One such alternative is
the Strongly Interacting Massive Particle (SIMP) which, by contrast, can have interactions with nucleons and electrons as strong as between these particles themselves~\cite{Choi:2017zww,Daci:2015hca,Albuquerque:2003ei,Mack:2007xj,Kouvaris:2014lpa,Hochberg:2014dra,Cudell:2014wca,Mahdawi:2017cxz}.
As can be expected, constraints on the interactions of SIMPs with the visible sector come from a wide variety of both terrestrial and astrophysical sources~\cite{RICH1987173,Starkman:1990nj,Natarajan:2002cw,Chen:2002yh,McDermott:2010pa,Daci:2015hca,Albuquerque:2003ei,Mack:2007xj,Kouvaris:2014lpa,Hochberg:2014dra,Cudell:2014wca,Foot:2014osa,Mahdawi:2017cxz}. However there remain values of the SIMP mass and interaction cross section
with the visible sector which have proven difficult to constrain in a model-independent way. One major reason for this is the ineffectiveness of direct searches in probing SIMP interactions. This results from the fact that SIMPs should scatter in
the Earth before reaching the experimental apparatus, causing them to lose kinetic energy such that by the time they reach the experiment they do not have enough energy to result in a nuclear recoil above the threshold~\cite{Albuquerque:2003ei,Kouvaris:2014lpa}.

In this work we discuss the potential for a surface-based direct dark matter search as a probe of the interaction cross section between SIMPs and nucleons. We re-analyse data from the recent dark matter search performed by the CRESST collaboration, described in refs.~\cite{Angloher:2017sxg,Strauss:2017cuu,Strauss:2017cam}, which was run on the surface of the Earth with only minimal shielding. This resulted in a large background rate making a WIMP search difficult, but the
low-threshold and lack of shielding makes such a set-up ideal for a SIMP search.

\section{Description of the experimental set-ups and data}
We consider two different scenarios for the low-threshold cryogenic experimental apparatus introduced in refs.~\cite{Angloher:2017sxg,Strauss:2017cuu,Strauss:2017cam}. 
For the experimental search for low-mass WIMP dark matter performed by the CRESST collaboration in ref.~\cite{Angloher:2017sxg}, which we refer to as the ``2017 Surface Run'', the experimental apparatus (named $\nu$-cleus) was run with only a small amount of shielding, which
amounted to  1mm of copper. This resulted in a large background which is detrimental to a standard WIMP search, but not for a SIMP search where the expected signal rates are much larger.

The apparatus was housed in a building at the Max-Planck-Institut for physics in Munich with walls of concrete approximately 30~cm thick, which actually provides the dominant stopping power for the SIMPs, besides the Earth and
atmosphere. The experiment was performed with a $0.49$~gram Al$_2$O$_3$ target running for a total live-time of 2.27~hours, with a low-energy threshold of $20$~eV. In this time, the experimental collaboration observed data consistent with their background expectation, which we take here to be a constant level of $10^5$ counts kg$^{-1}$ keV$^{-1}$ day$^{-1}$~\cite{Angloher:2017sxg}.

We also consider a future scenario for such a detector with as low a threshold as possible for nuclear recoils. This scenario, which we call the ``Ideal Surface Projection'' would utilise a cryogenic set-up with an Al$_2$O$_3$ target, with
a low-energy threshold of 4~eV. 
Under this scenario we also make the assumption that the 30~cm of concrete shielding has been removed, however the effect of this on the sensitivity of the experimental appratus to large SIMP-nucleus cross sections
is likely to be minimal, compared with the improvement gained through lowering the threshold. This is because even with 30~cm of concerete shielding the dominant energy loss mechanism for the SIMP particles will arise through scattering
while travelling through the Earth's atmosphere. Hence it is possible that the CRESST collaboration may find it more prudent to increase their shielding enough to dramatically improve their sensitivity to small DM-nucleon cross sections,
while only slightly reducing their sentivity to larger cross sections.

\section{Calculating the velocity distribution of SIMPs at the detector}
In this section we present analytic calculations of the energy loss experienced by SIMPs as they travel through the Earth, the atmosphere and the shielding around the experiment.
We want to know what the distribution of the SIMP kinetic energies will be at the detector, given that we know this distribution in free space.
We start by assuming a model for the distribution of SIMPs in the galactic halo, which we take to be the same as for WIMPs~\cite{Angloher:2017sxg}.
Hence we sample the initial SIMP energies $E_i$ from a Maxwell-Boltzmann velocity distribution with a maximum velocity equal to the galactic escape velocity, which has been boosted into the Earth's reference frame. For each $E_i$ we then calculate the final energy $E_f$ using the following process.

We assume that SIMPs interact only with
nuclei via spin-independent (SI) contact interactions, and hence the differential scattering cross section of SIMPs with mass $m_{\chi}$ interacting with nuclei of mass $m_N$ takes the form,
\begin{equation}
\frac{\mathrm{d}\sigma}{\mathrm{d}E_R} = \frac{m_N \sigma_n A^2}{2 \mu_p^2 v^2},
\label{eqn:dsigmader}
\end{equation}
where $E_R$ is the recoil energy transferred from the SIMPs, $\sigma_n$ is the SIMP-nucleon cross section, $A$ is the atomic mass of the stopping nucleus, $v$ is the velocity of the SIMP particle and $\mu_p$ is the SIMP-proton reduced
mass.

The stopping power for SIMPs passing through a material composed of a single element is then calculated using~\cite{Kouvaris:2014lpa,Emken:2017erx},
\begin{eqnarray}
\frac{\mathrm{d}E}{\mathrm{d}x} = -n_N \int_0^{E_{R,\mathrm{max}}} \frac{\mathrm{d}\sigma}{\mathrm{d}E_R} E_R \mathrm{d} E_R \\
E_{R,\mathrm{max}} = \frac{4 m_{\chi} m_N E}{ (m_{\chi} + m_N)^2 },
\end{eqnarray}
which when combined with equation (\ref{eqn:dsigmader}) and integrating over $E_R$ gives,
\begin{equation}
E_f = E_i \exp \left[ - \int_0^{\infty} \frac{2 \sigma_n A^2 \mu_N^4 n_N(l)}{m_{\chi} m_N \mu_p^2} \, \mathrm{d}l \right] ,
\label{eqn:energy_loss_1}
\end{equation}
where $\mu_N$ is the SIMP-nucleus reduced mass, $n_N(l)$ is the density of the stopping matter and we have changed variables from $x$ to $l$, the distance between the SIMP particle and the detector.

In reality the SIMP will pass through material composed of multiple elements, and so we will have to generalise equation~(\ref{eqn:energy_loss_1}) to multiple targets.
Indeed the density function $n_N(l)$ takes a different form depending on whether a SIMP is travelling through the Earth, the atmosphere or the sheilding around the experiment, as does the elemental composition of the target.
Furthermore in the first two cases, the density depends on 
the radial distance from the centre of the Earth to the SIMP $r$.
In order to relate the distance from the centre of the Earth to the SIMP $r$ to the distance of the SIMP from the detector $l$ we use the expression~\cite{Kouvaris:2014lpa,Emken:2017erx},
\begin{equation}
r^2 = (R_E - l_D)^2 + l^2 - 2(R_E - l_D) l \cos{\psi} ,
\label{eqn:radial_distance}
\end{equation}
where $R_E$ is the radius of the Earth and $l_D$ is the depth of the experiment, where $l_D = 0$ for a surface-based detector. The angle $\psi$ is between the direction of the vector pointing along the travel direction of the SIMP
towards the detector, and the vector between the centre of the Earth and the detector~\cite{Kouvaris:2014lpa,Emken:2017erx}. It is expressed as,
\begin{equation}
\begin{split}
 \cos{\psi} = \cos \theta_l \cos \omega t \sin \theta \cos \phi + \\ \cos \theta_l \sin \omega t \sin \theta \sin \phi \pm \sin \theta_l \cos \theta ,
\end{split}
 \label{eqn:psi}
\end{equation}
where $\omega$ is the angular rotation speed of the Earth, $t$ is time, $\theta_l$ is the latitude of the detector and the $\pm$ is $+$ for the northern hemisphere and $-$ for the southern.

For clarity we split the expression
of equation (\ref{eqn:energy_loss_1}) into three pieces i.e. an integral over all values of $l$ which fall within either the Earth, the atmosphere or the shielding, and express all of the target-dependent terms as a single variable
$\mathcal{N}(x) = \sum_i f_i A_i^2 \mu_{N,i}^4 n_N(x) / m_{N,i}$, where $i$ is summed over all constituent elements of the stopping target with mass-fraction $f_i$. Hence the expression for $E_f$ becomes,
\begin{equation}
\begin{split}
E_f = E_i \exp \Bigg[ -  \frac{2 \sigma_n}{m_{\chi} \mu_p^2} \Bigg(  \int\limits_{\mathrm{Atmosphere}}  \mathcal{N}_A(r-R_E)  \, \mathrm{d}l  + \\
 \int\limits_{\mathrm{Earth}}  \mathcal{N}_E(r)  \, \mathrm{d}l + 
 \int\limits_{\mathrm{Sheilding}}  \mathcal{N}_S(l)  \, \mathrm{d}l
\Bigg) \Bigg] ,
\end{split}
\end{equation}
where $\mathcal{N}_E(r)$ and $ \mathcal{N}_A(r-R_E) $ contain all of the target-dependent terms from equation~(\ref{eqn:energy_loss_1}) for scattering in either the Earth or the atmosphere respectively, and are the same for each 
experimental set-up at a given latitude and depth. The term $\mathcal{N}_S(l)$ contains all target-dependent terms concerning the experimental shielding.

To calculate the Earth-stopping term $\mathcal{N}_E(r)$ we use the Preliminary Reference Earth Model~\cite{DZIEWONSKI1981297,Emken:2017qmp} for both the elemental abundances and the radial dependence of the stopping target density.
The term $ \mathcal{N}_A(r-R_E) $ for the stopping power of the atmosphere for SIMPs depends on the height above the Earth's surface $h = r - R_E$. For this we use the US Standard Atmosphere 1976 model~\cite{US_atmosphere}.
After integrating over $l$ for each sampled initial SIMP kinetic energy $E_i$ we have a distribution of $E_f$ values, which we then use to calculate a normalised histogram for the distribution of SIMP speeds $f(v_f)$.

It is important to understand the limitations of our analytic treatment in calculating the spectra of SIMPs. In particular our simplifying assumptions will introduce a factor of a few uncertainty to the largest values of the SIMP-nucleon cross section which can be excluded
by $\nu$-cleus.
Indeed a numerical code DaMaSCUS~\cite{Emken:2017qmp}
exists for calculating the effect of Earth stopping on SIMPs for underground experiments. However since we are considering a surface-based experiment in this work, for which scattering in the atmosphere and shielding dominates the sensitivity
to SIMPs, we can not make effective use of this code. A comparison of the analytic and numerical methods for SIMP stopping was made in ref.~\cite{Emken:2017erx}, where it was found that the maximum excluded SIMP cross section was
around an order of magnitude smaller using the numerical code, compared to an analytic calculation. This was due mainly to the fact that the path length travelled by the SIMPs in the Earth is lengthened due to deflection of the particles
through scattering. 

In addition, our calculations focused only on the average energy loss rate per SIMP, whereas it was pointed out in ref.~\cite{Mahdawi:2017cxz} that if a SIMP particle were to trigger a signal in a direct detection
experiment, it would likely be sampled from the tail-end of the statistical distribution in path length and the energy-loss per scatter. Hence even if the average energy of SIMPs is too low to induce a signal in an experiment above threshold, there could still be 
such out-lying SIMPs which could lead to a signal. The effect of this would be to work in the opposite way to the path-lenghtening effect, strengthening the ability of experiments such as $\nu$-cleus to exclude larger SIMP-nucleon cross
sections~\cite{Mahdawi:2017cxz}. Hence the order-of-magnitude effect in ref.~\cite{Emken:2017erx} is likely an upper bound on the problem.

\section{Recoil spectra of SIMPs}
\begin{figure}[t]
\includegraphics[width=0.47\textwidth]{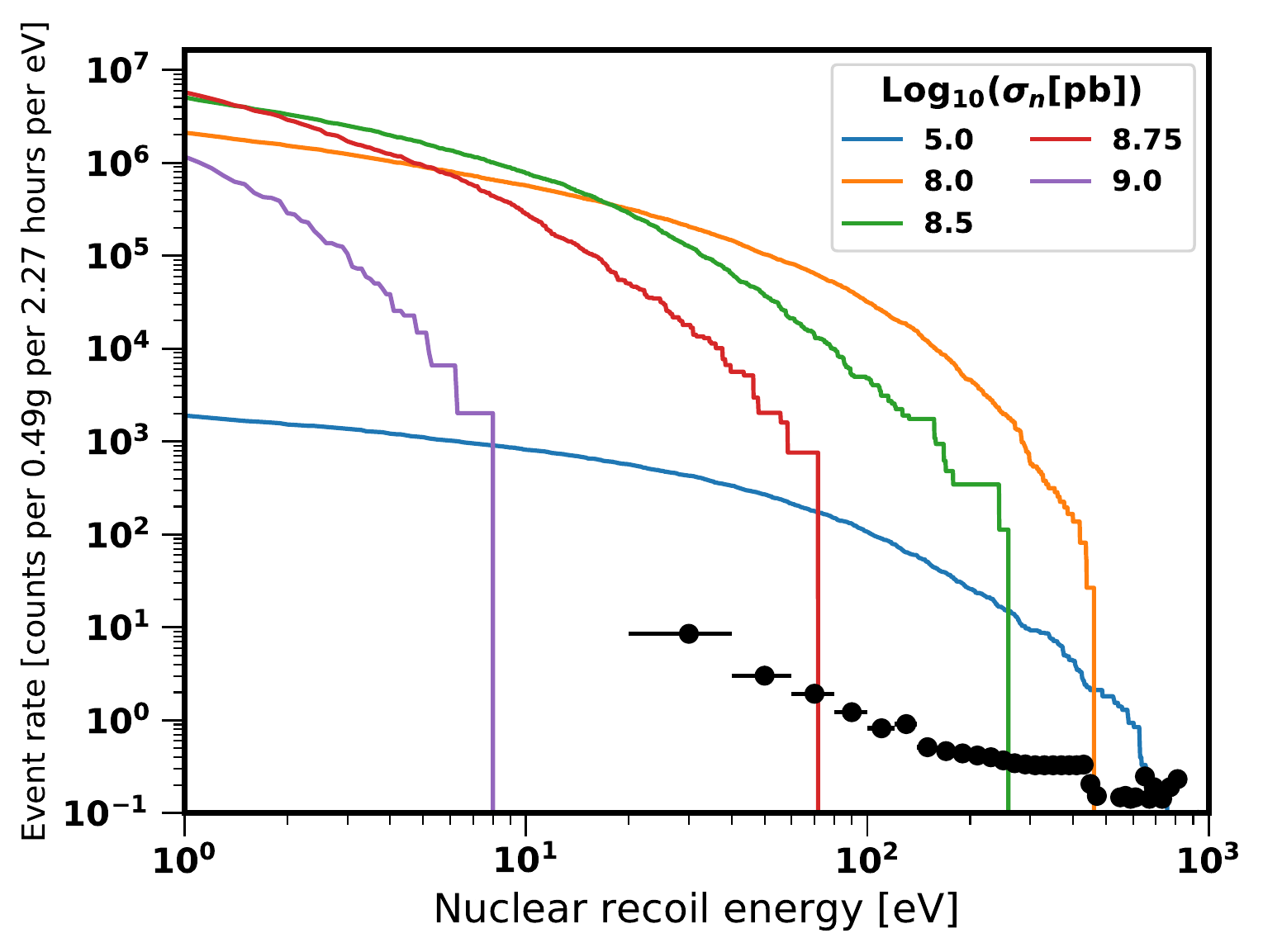}
\caption{Recoil spectra of 1~GeV mass SIMP dark matter for different values of the interaction cross section with nucleons $\sigma_n$, averaged over the duration of the experimental run time, for the ``2017 Surface Run'' configuration.
We also show as black points the data from the dark matter search performed by the CRESST collaboration using the $\nu$-cleus experiment in ref.~\cite{Angloher:2017sxg}, which we use to set exclusion limits in this work.}
\label{fig:spectra_1}
\end{figure}

Using the distribution of SIMP speeds reaching the detector, which we calculated in the previous section, it is possible to determine the spectrum of nuclear recoil energies $E_r$ in the experiment due to SIMPs. 
Since in our case the experimental apparatus has no sensitivity to the SIMP incident direction, we integrate over all arrival angles, leading to the formula for the recoil spectrum,
\begin{equation}
\frac{\mathrm{d} R}{\mathrm{d} E_r} = \frac{\rho_{\chi}}{m_{\chi}} \sum_i f_i N_T \int_{v_{\mathrm{min}}} \frac{\mathrm{d} \sigma}{\mathrm{d} E_r} f(v_f) \, \mathrm{d} v_f,
\label{eqn:drde}
\end{equation}
where $N_T$ is the total number of target nuclei in the detector, $v_{\mathrm{min}} = \sqrt{E_r m_N / 2 \mu_N^2}$, the minimum SIMP speed needed to impart a recoil energy $E_r$ to a nucleus with mass $m_N$, and the sum is over all elements which make up the experimental target with mass fraction
$f_i$.

This recoil spectrum varies over a period of a day due to the time-dependence of the angle $\psi$ in equation (\ref{eqn:psi}), combined with the fact that dark matter particles arrive at Earth from a preferred direction~\cite{Kouvaris:2014lpa}. Hence at certain points in the day more SIMPs have to traverse a longer distance through the Earth than at other times, leading to enhanced stopping and deflection of particles~\cite{Emken:2017erx,Kavanagh:2016pyr}. We do not consider the latter effect in this work, as we are only interested in order-of-magnitude estimates for the sensitivity of surface-based detectors to SIMPs.

Shown in figure~\ref{fig:spectra_1} are the time-averaged recoil spectra for a 1~GeV mass SIMP and various different values of $\sigma_n$. With increasing cross section the rate of events in the detector
increases up until a certain value, at which point the SIMPs scatter enough in the Earth, the atmosphere or the experimental shielding to undergo significant energy loss before they reach the detector. This results in the
spectrum shifting to smaller values of recoil energy, eventually leading to all of the nuclear recoils occurring below the experimental threshold, making the SIMP scatter events invisible to the detector.
Hence one advantage of having a lower experimental threshold, besides improved sensitivity to lower mass particles, is a greater sensitivity to larger values SIMP-nucleon cross section. For example a 1~GeV mass SIMP with a cross section 
of $\sigma_n = 10^9$~pb would be invisible to the 2017 Surface Run but would be just detectable using the lower threshold for the Ideal Surface Projection.

All of the SIMP recoil rates shown in figure~\ref{fig:spectra_1} are well above the rate observed in the surface run of ref.~\cite{Angloher:2017sxg}, shown as black data-points in figure~\ref{fig:spectra_1}, which levels out to 
around $10^5$ counts kg$^{-1}$ keV$^{-1}$ day$^{-1}$ at higher energies.
Hence, although such a large background is detrimental to a standard
WIMP dark matter search, it has only a small effect on the sensitivity to high SIMP cross sections. 
For our analysis we assume a constant background rate, despite the fact that the data from the $\nu$-cleus experiment (shown in figure~\ref{fig:spectra_1}) rises at low energies. This allows us to place a conservative limit
on the excluded region of SIMP parameter space, given that the background near the low-energy threshold is not perfectly understood. The data is assumed to originate entirely from backgrounds, and not DM-nucleon scattering events.

The expected rate of interactions is so large that individual nuclear recoils may not be resolvable in the $\nu$-cleus apparatus, appearing instead as a uniform heating~\cite{Strauss:2017cuu}. Under a conservative estimate of a pulse-separation resolution of $1$~ms the maximum observable rate would be approximately $\sim 10^7$~events per day. However for a resolution of $10$~$\mu$s~\cite{Strauss:2017cuu}
the maximum observable rate of individual recoil events would be closer to $\sim 10^9$~events per day, and so all of the cross sections considered in figure~\ref{fig:spectra_1} would lead to observable spectra in $\nu$-cleus.

\section{Results}
\begin{figure}[t]
\includegraphics[width=0.47\textwidth]{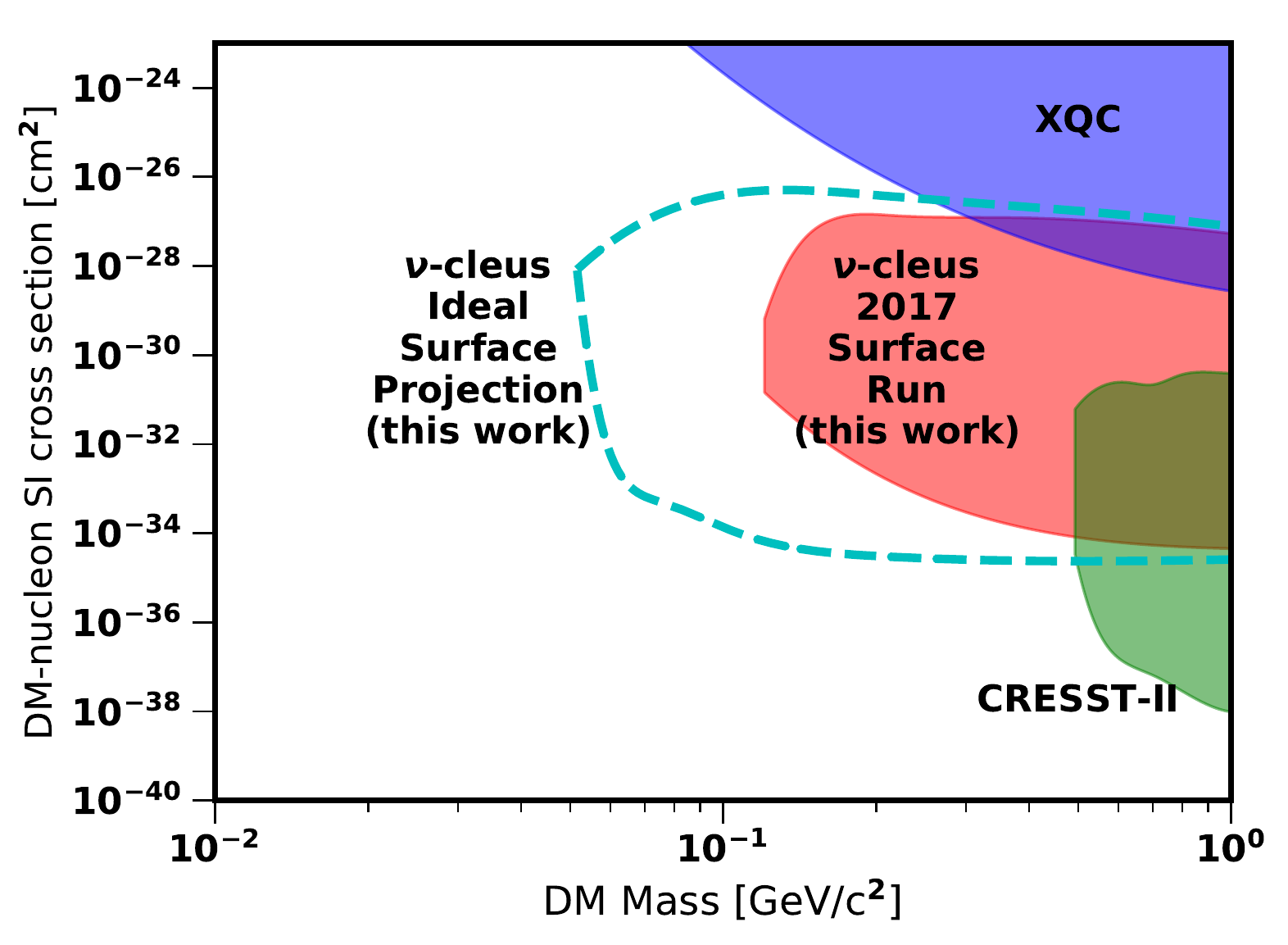}
\caption{Excluded regions of dark matter (DM) mass and cross section at $95\%$ confidence from various ``direct'' experiments (filled), including our reanalysis of the data from the dark matter search performed by the CRESST collaboration with the $\nu$-cleus apparatus in ref.~\cite{Angloher:2017sxg}, and the  projected region of exclusion (dashed line) from a surface run using the apparatus described in this work with a low-energy
threshold of 4~eV.}
\label{fig:simp_results}
\end{figure}

Using equation (\ref{eqn:drde}) we have performed a likelihood-ratio parameter scan over the SIMP cross section for various different masses using the data from the 2017 Surface Run of the $\nu$-cleus experiment, obtained
by the CRESST collaboration in ref.~\cite{Angloher:2017sxg}. Since no excess over background expectation was observed we have derived an exclusion region at $95\%$ confidence. We have also performed a similar analysis on simulated data for the Ideal Surface Projection under the assumption that, were such an experiment performed, that the data would be consistent with a background expectation, with the same background rate.

In figure~\ref{fig:simp_results} we show the $95\%$ confidence excluded region set using the data from ref.~\cite{Angloher:2017sxg} i.e. the 2017 Surface Run, and our Ideal Surface Projection with a 4eV threshold. This is compared with excluded regions from the CRESST-II dark matter
search experiment~\cite{Angloher:2015ewa}, which is 1.4~km underground, and
the X-ray Quantum Calorimetry Experiment (XQC)~\cite{Erickcek:2007jv}, which was launched on a rocket up to an altitude of 225~km above the Earth's surface, thereby considerably reducing the stopping from the atmosphere, but at the
expense of a huge background rate for a dark matter search. Our excluded region complements both high-altitude searches such as XQC and underground searches such as CRESST-II, whose sensitivity to cross sections above around
$10^{-31}$~cm$^2$ is limited by scattering of the SIMPs in the rock over-burden.

The main improvement which would be gained using the Ideal Surface Projection set-up is an improved sensitivity to lower masses, but without much gain towards higher cross sections at larger masses around 1~GeV. As expected, the limiting
factor in the sensitivity of the 2017 Surface Run to SIMPs is not the stopping power from the $\sim 30$~cm concrete but actually the Earth's atmosphere, and the dominant factor in the improvement gained with the ideal projection is the lower threshold. Hence the results presented here are likely to be the strongest sensitivity which
can be achieved to large cross sections with an experiment based on the surface of the Earth.

Beyond direct constraints, there are also complimentary limits on SIMP dark matter from collider searches~\cite{Daci:2015hca}, searches for new forces between nuclei~\cite{Fichet:2017bng}, astrophysical observations such as neutron stars~\cite{Baryakhtar:2017dbj,Kouvaris:2010vv}, large-scale structure and the Cosmic Microwave Background~\cite{Chen:2002yh} or the heat budget of the Earth~\cite{Mack:2007xj}.

\section{Conclusion}
Although most dark matter searches focus on weakly interacting particles e.g. WIMPs, it is worthwhile to consider alternatives such as dark matter which interacts more readily with nucleons.
Searching for such SIMP dark matter with underground direct detection experiments is difficult, since the SIMPs lose a significant amount of their kinetic energy travelling through the rock overburden~\cite{Albuquerque:2003ei,Emken:2017qmp,Kouvaris:2014lpa,Mack:2007xj,Mahdawi:2017cxz}. 
Hence in order to maximise sensitivity to SIMPs, a direct detection experiment would need to be based on the surface of the Earth with minimal shielding. Though even in this case, as shown in figure~\ref{fig:spectra_1}, the SIMP spectrum is still 
pushed below threshold for high enough SIMP-nucleon cross sections, mainly due to energy loss of the SIMPs in the Earth's atmosphere. 
 
A surface run of a low-threshold cryogenic dark matter direct detection experiment $\nu$-cleus, performed by the CRESST collaboration in ref.~\cite{Angloher:2017sxg}, found no evidence for an excess of events above background expectation. In this work we have re-analysed this data in the context of SIMP dark matter, to place constraints on the scattering cross section between SIMPs lighter than a GeV and nucleons. As shown in figure~\ref{fig:simp_results} the constraint from this search opens up a new region of parameter space bounded from below by underground direct dark matter searches~\cite{Angloher:2015ewa} and from above by high-altitude experiments such as a the rocket-bourne X-ray Quantum Calorimetry Experiment~\cite{Erickcek:2007jv}.
Furthermore by reducing the low-energy nuclear recoil  threshold to an experimentally viable target of 4~eV the SIMP parameter space can be probed down to even smaller masses, as low as 60~MeV.
 
\begin{acknowledgments} 
The author thanks Raimund Strauss for information about the set-up of the $\nu$-cleus experiment, and low-threshold cryogenic detectors in general, and Florian Reindl for comments on the manuscript.
The research leading to these results has received funding from the European Research Council through the project DARKHORIZONS under the European Union's Horizon 2020 program (ERC Grant Agreement no.648680). 
\end{acknowledgments}  


\begin{thebibliography}{38}%
\makeatletter
\providecommand \@ifxundefined [1]{%
 \@ifx{#1\undefined}
}%
\providecommand \@ifnum [1]{%
 \ifnum #1\expandafter \@firstoftwo
 \else \expandafter \@secondoftwo
 \fi
}%
\providecommand \@ifx [1]{%
 \ifx #1\expandafter \@firstoftwo
 \else \expandafter \@secondoftwo
 \fi
}%
\providecommand \natexlab [1]{#1}%
\providecommand \enquote  [1]{``#1''}%
\providecommand \bibnamefont  [1]{#1}%
\providecommand \bibfnamefont [1]{#1}%
\providecommand \citenamefont [1]{#1}%
\providecommand \href@noop [0]{\@secondoftwo}%
\providecommand \href [0]{\begingroup \@sanitize@url \@href}%
\providecommand \@href[1]{\@@startlink{#1}\@@href}%
\providecommand \@@href[1]{\endgroup#1\@@endlink}%
\providecommand \@sanitize@url [0]{\catcode `\\12\catcode `\$12\catcode
  `\&12\catcode `\#12\catcode `\^12\catcode `\_12\catcode `\%12\relax}%
\providecommand \@@startlink[1]{}%
\providecommand \@@endlink[0]{}%
\providecommand \url  [0]{\begingroup\@sanitize@url \@url }%
\providecommand \@url [1]{\endgroup\@href {#1}{\urlprefix }}%
\providecommand \urlprefix  [0]{URL }%
\providecommand \Eprint [0]{\href }%
\providecommand \doibase [0]{http://dx.doi.org/}%
\providecommand \selectlanguage [0]{\@gobble}%
\providecommand \bibinfo  [0]{\@secondoftwo}%
\providecommand \bibfield  [0]{\@secondoftwo}%
\providecommand \translation [1]{[#1]}%
\providecommand \BibitemOpen [0]{}%
\providecommand \bibitemStop [0]{}%
\providecommand \bibitemNoStop [0]{.\EOS\space}%
\providecommand \EOS [0]{\spacefactor3000\relax}%
\providecommand \BibitemShut  [1]{\csname bibitem#1\endcsname}%
\let\auto@bib@innerbib\@empty
\bibitem [{\citenamefont {Ade}\ \emph {et~al.}(2016)\citenamefont {Ade} \emph
  {et~al.}}]{Ade:2015xua}%
  \BibitemOpen
  \bibfield  {author} {\bibinfo {author} {\bibfnamefont {P.~A.~R.}\
  \bibnamefont {Ade}} \emph {et~al.} (\bibinfo {collaboration} {Planck}),\
  }\bibfield  {title} {\enquote {\bibinfo {title} {{Planck 2015 results. XIII.
  Cosmological parameters}},}\ }\href {\doibase 10.1051/0004-6361/201525830}
  {\bibfield  {journal} {\bibinfo  {journal} {Astron. Astrophys.}\ }\textbf
  {\bibinfo {volume} {594}},\ \bibinfo {pages} {A13} (\bibinfo {year}
  {2016})},\ \Eprint {http://arxiv.org/abs/1502.01589} {arXiv:1502.01589
  [astro-ph.CO]} \BibitemShut {NoStop}%
\bibitem [{\citenamefont {Davis}\ and\ \citenamefont
  {Silk}(2015)}]{Davis:2014zda}%
  \BibitemOpen
  \bibfield  {author} {\bibinfo {author} {\bibfnamefont {Jonathan~H.}\
  \bibnamefont {Davis}}\ and\ \bibinfo {author} {\bibfnamefont {Joseph}\
  \bibnamefont {Silk}},\ }\bibfield  {title} {\enquote {\bibinfo {title} {{Glow
  in the Dark Matter: Observing galactic halos with scattered light}},}\ }\href
  {\doibase 10.1103/PhysRevLett.114.051303} {\bibfield  {journal} {\bibinfo
  {journal} {Phys. Rev. Lett.}\ }\textbf {\bibinfo {volume} {114}},\ \bibinfo
  {pages} {051303} (\bibinfo {year} {2015})},\ \Eprint
  {http://arxiv.org/abs/1410.5423} {arXiv:1410.5423 [hep-ph]} \BibitemShut
  {NoStop}%
\bibitem [{\citenamefont {Aprile}\ \emph {et~al.}(2017)\citenamefont {Aprile}
  \emph {et~al.}}]{Aprile:2017iyp}%
  \BibitemOpen
  \bibfield  {author} {\bibinfo {author} {\bibfnamefont {E.}~\bibnamefont
  {Aprile}} \emph {et~al.} (\bibinfo {collaboration} {XENON}),\ }\bibfield
  {title} {\enquote {\bibinfo {title} {{First Dark Matter Search Results from
  the XENON1T Experiment}},}\ }\href@noop {} {\  (\bibinfo {year} {2017})},\
  \Eprint {http://arxiv.org/abs/1705.06655} {arXiv:1705.06655 [astro-ph.CO]}
  \BibitemShut {NoStop}%
\bibitem [{\citenamefont {Akerib}\ \emph {et~al.}(2017)\citenamefont {Akerib}
  \emph {et~al.}}]{Akerib:2016vxi}%
  \BibitemOpen
  \bibfield  {author} {\bibinfo {author} {\bibfnamefont {D.~S.}\ \bibnamefont
  {Akerib}} \emph {et~al.} (\bibinfo {collaboration} {LUX}),\ }\bibfield
  {title} {\enquote {\bibinfo {title} {{Results from a search for dark matter
  in the complete LUX exposure}},}\ }\href {\doibase
  10.1103/PhysRevLett.118.021303} {\bibfield  {journal} {\bibinfo  {journal}
  {Phys. Rev. Lett.}\ }\textbf {\bibinfo {volume} {118}},\ \bibinfo {pages}
  {021303} (\bibinfo {year} {2017})},\ \Eprint
  {http://arxiv.org/abs/1608.07648} {arXiv:1608.07648 [astro-ph.CO]}
  \BibitemShut {NoStop}%
\bibitem [{\citenamefont {Agnese}\ \emph {et~al.}(2016)\citenamefont {Agnese}
  \emph {et~al.}}]{Agnese:2015nto}%
  \BibitemOpen
  \bibfield  {author} {\bibinfo {author} {\bibfnamefont {R.}~\bibnamefont
  {Agnese}} \emph {et~al.} (\bibinfo {collaboration} {SuperCDMS}),\ }\bibfield
  {title} {\enquote {\bibinfo {title} {{New Results from the Search for
  Low-Mass Weakly Interacting Massive Particles with the CDMS Low Ionization
  Threshold Experiment}},}\ }\href {\doibase 10.1103/PhysRevLett.116.071301}
  {\bibfield  {journal} {\bibinfo  {journal} {Phys. Rev. Lett.}\ }\textbf
  {\bibinfo {volume} {116}},\ \bibinfo {pages} {071301} (\bibinfo {year}
  {2016})},\ \Eprint {http://arxiv.org/abs/1509.02448} {arXiv:1509.02448
  [astro-ph.CO]} \BibitemShut {NoStop}%
\bibitem [{\citenamefont {Tan}\ \emph {et~al.}(2016)\citenamefont {Tan} \emph
  {et~al.}}]{Tan:2016zwf}%
  \BibitemOpen
  \bibfield  {author} {\bibinfo {author} {\bibfnamefont {Andi}\ \bibnamefont
  {Tan}} \emph {et~al.} (\bibinfo {collaboration} {PandaX-II}),\ }\bibfield
  {title} {\enquote {\bibinfo {title} {{Dark Matter Results from First 98.7
  Days of Data from the PandaX-II Experiment}},}\ }\href {\doibase
  10.1103/PhysRevLett.117.121303} {\bibfield  {journal} {\bibinfo  {journal}
  {Phys. Rev. Lett.}\ }\textbf {\bibinfo {volume} {117}},\ \bibinfo {pages}
  {121303} (\bibinfo {year} {2016})},\ \Eprint
  {http://arxiv.org/abs/1607.07400} {arXiv:1607.07400 [hep-ex]} \BibitemShut
  {NoStop}%
\bibitem [{\citenamefont {Knapen}\ \emph {et~al.}(2017)\citenamefont {Knapen},
  \citenamefont {Lin},\ and\ \citenamefont {Zurek}}]{Knapen:2016cue}%
  \BibitemOpen
  \bibfield  {author} {\bibinfo {author} {\bibfnamefont {Simon}\ \bibnamefont
  {Knapen}}, \bibinfo {author} {\bibfnamefont {Tongyan}\ \bibnamefont {Lin}}, \
  and\ \bibinfo {author} {\bibfnamefont {Kathryn~M.}\ \bibnamefont {Zurek}},\
  }\bibfield  {title} {\enquote {\bibinfo {title} {{Light Dark Matter in
  Superfluid Helium: Detection with Multi-excitation Production}},}\ }\href
  {\doibase 10.1103/PhysRevD.95.056019} {\bibfield  {journal} {\bibinfo
  {journal} {Phys. Rev.}\ }\textbf {\bibinfo {volume} {D95}},\ \bibinfo {pages}
  {056019} (\bibinfo {year} {2017})},\ \Eprint
  {http://arxiv.org/abs/1611.06228} {arXiv:1611.06228 [hep-ph]} \BibitemShut
  {NoStop}%
\bibitem [{\citenamefont {Angloher}\ \emph {et~al.}(2017)\citenamefont
  {Angloher} \emph {et~al.}}]{Angloher:2017sxg}%
  \BibitemOpen
  \bibfield  {author} {\bibinfo {author} {\bibfnamefont {G.}~\bibnamefont
  {Angloher}} \emph {et~al.} (\bibinfo {collaboration} {CRESST}),\ }\bibfield
  {title} {\enquote {\bibinfo {title} {{Results on MeV-scale dark matter from a
  gram-scale cryogenic calorimeter operated above ground}},}\ }\href@noop {} {\
   (\bibinfo {year} {2017})},\ \Eprint {http://arxiv.org/abs/1707.06749}
  {arXiv:1707.06749 [astro-ph.CO]} \BibitemShut {NoStop}%
\bibitem [{\citenamefont {Strauss}\ \emph
  {et~al.}(2017{\natexlab{a}})\citenamefont {Strauss} \emph
  {et~al.}}]{Strauss:2017cuu}%
  \BibitemOpen
  \bibfield  {author} {\bibinfo {author} {\bibfnamefont {R.}~\bibnamefont
  {Strauss}} \emph {et~al.},\ }\bibfield  {title} {\enquote {\bibinfo {title}
  {{The $\nu$-cleus experiment: A gram-scale fiducial-volume cryogenic detector
  for the first detection of coherent neutrino-nucleus scattering}},}\ }\href
  {\doibase 10.1140/epjc/s10052-017-5068-2} {\bibfield  {journal} {\bibinfo
  {journal} {Eur. Phys. J.}\ }\textbf {\bibinfo {volume} {C77}},\ \bibinfo
  {pages} {506} (\bibinfo {year} {2017}{\natexlab{a}})},\ \Eprint
  {http://arxiv.org/abs/1704.04320} {arXiv:1704.04320 [physics.ins-det]}
  \BibitemShut {NoStop}%
\bibitem [{\citenamefont {Strauss}\ \emph
  {et~al.}(2017{\natexlab{b}})\citenamefont {Strauss} \emph
  {et~al.}}]{Strauss:2017cam}%
  \BibitemOpen
  \bibfield  {author} {\bibinfo {author} {\bibfnamefont {R.}~\bibnamefont
  {Strauss}} \emph {et~al.},\ }\bibfield  {title} {\enquote {\bibinfo {title}
  {{Gram-scale cryogenic calorimeters for rare-event searches}},}\ }\href
  {\doibase 10.1103/PhysRevD.96.022009} {\bibfield  {journal} {\bibinfo
  {journal} {Phys. Rev.}\ }\textbf {\bibinfo {volume} {D96}},\ \bibinfo {pages}
  {022009} (\bibinfo {year} {2017}{\natexlab{b}})},\ \Eprint
  {http://arxiv.org/abs/1704.04317} {arXiv:1704.04317 [physics.ins-det]}
  \BibitemShut {NoStop}%
\bibitem [{\citenamefont {Angloher}\ \emph {et~al.}(2016)\citenamefont
  {Angloher} \emph {et~al.}}]{Angloher:2015ewa}%
  \BibitemOpen
  \bibfield  {author} {\bibinfo {author} {\bibfnamefont {G.}~\bibnamefont
  {Angloher}} \emph {et~al.} (\bibinfo {collaboration} {CRESST}),\ }\bibfield
  {title} {\enquote {\bibinfo {title} {{Results on light dark matter particles
  with a low-threshold CRESST-II detector}},}\ }\href {\doibase
  10.1140/epjc/s10052-016-3877-3} {\bibfield  {journal} {\bibinfo  {journal}
  {Eur. Phys. J.}\ }\textbf {\bibinfo {volume} {C76}},\ \bibinfo {pages} {25}
  (\bibinfo {year} {2016})},\ \Eprint {http://arxiv.org/abs/1509.01515}
  {arXiv:1509.01515 [astro-ph.CO]} \BibitemShut {NoStop}%
\bibitem [{\citenamefont {McCabe}(2017)}]{McCabe:2017rln}%
  \BibitemOpen
  \bibfield  {author} {\bibinfo {author} {\bibfnamefont {Christopher}\
  \bibnamefont {McCabe}},\ }\bibfield  {title} {\enquote {\bibinfo {title}
  {{New constraints and discovery potential of sub-GeV dark matter with xenon
  detectors}},}\ }\href@noop {} {\  (\bibinfo {year} {2017})},\ \Eprint
  {http://arxiv.org/abs/1702.04730} {arXiv:1702.04730 [hep-ph]} \BibitemShut
  {NoStop}%
\bibitem [{\citenamefont {Kouvaris}\ and\ \citenamefont
  {Pradler}(2017)}]{Kouvaris:2016afs}%
  \BibitemOpen
  \bibfield  {author} {\bibinfo {author} {\bibfnamefont {Chris}\ \bibnamefont
  {Kouvaris}}\ and\ \bibinfo {author} {\bibfnamefont {Josef}\ \bibnamefont
  {Pradler}},\ }\bibfield  {title} {\enquote {\bibinfo {title} {{Probing
  sub-GeV Dark Matter with conventional detectors}},}\ }\href {\doibase
  10.1103/PhysRevLett.118.031803} {\bibfield  {journal} {\bibinfo  {journal}
  {Phys. Rev. Lett.}\ }\textbf {\bibinfo {volume} {118}},\ \bibinfo {pages}
  {031803} (\bibinfo {year} {2017})},\ \Eprint
  {http://arxiv.org/abs/1607.01789} {arXiv:1607.01789 [hep-ph]} \BibitemShut
  {NoStop}%
\bibitem [{\citenamefont {Agnese}\ \emph {et~al.}(2017)\citenamefont {Agnese}
  \emph {et~al.}}]{Agnese:2017jvy}%
  \BibitemOpen
  \bibfield  {author} {\bibinfo {author} {\bibfnamefont {R.}~\bibnamefont
  {Agnese}} \emph {et~al.} (\bibinfo {collaboration} {SuperCDMS}),\ }\bibfield
  {title} {\enquote {\bibinfo {title} {{Low-Mass Dark Matter Search with
  CDMSlite}},}\ }\href@noop {} {\bibfield  {journal} {\bibinfo  {journal}
  {Submitted to: Phys. Rev. D}\ } (\bibinfo {year} {2017})},\ \Eprint
  {http://arxiv.org/abs/1707.01632} {arXiv:1707.01632 [astro-ph.CO]}
  \BibitemShut {NoStop}%
\bibitem [{\citenamefont {Hochberg}\ \emph {et~al.}(2017)\citenamefont
  {Hochberg}, \citenamefont {Lin},\ and\ \citenamefont
  {Zurek}}]{Hochberg:2016sqx}%
  \BibitemOpen
  \bibfield  {author} {\bibinfo {author} {\bibfnamefont {Yonit}\ \bibnamefont
  {Hochberg}}, \bibinfo {author} {\bibfnamefont {Tongyan}\ \bibnamefont {Lin}},
  \ and\ \bibinfo {author} {\bibfnamefont {Kathryn~M.}\ \bibnamefont {Zurek}},\
  }\bibfield  {title} {\enquote {\bibinfo {title} {{Absorption of light dark
  matter in semiconductors}},}\ }\href {\doibase 10.1103/PhysRevD.95.023013}
  {\bibfield  {journal} {\bibinfo  {journal} {Phys. Rev.}\ }\textbf {\bibinfo
  {volume} {D95}},\ \bibinfo {pages} {023013} (\bibinfo {year} {2017})},\
  \Eprint {http://arxiv.org/abs/1608.01994} {arXiv:1608.01994 [hep-ph]}
  \BibitemShut {NoStop}%
\bibitem [{\citenamefont {Choi}\ \emph {et~al.}(2017)\citenamefont {Choi},
  \citenamefont {Hochberg}, \citenamefont {Kuflik}, \citenamefont {Lee},
  \citenamefont {Mambrini}, \citenamefont {Murayama},\ and\ \citenamefont
  {Pierre}}]{Choi:2017zww}%
  \BibitemOpen
  \bibfield  {author} {\bibinfo {author} {\bibfnamefont {Soo-Min}\ \bibnamefont
  {Choi}}, \bibinfo {author} {\bibfnamefont {Yonit}\ \bibnamefont {Hochberg}},
  \bibinfo {author} {\bibfnamefont {Eric}\ \bibnamefont {Kuflik}}, \bibinfo
  {author} {\bibfnamefont {Hyun~Min}\ \bibnamefont {Lee}}, \bibinfo {author}
  {\bibfnamefont {Yann}\ \bibnamefont {Mambrini}}, \bibinfo {author}
  {\bibfnamefont {Hitoshi}\ \bibnamefont {Murayama}}, \ and\ \bibinfo {author}
  {\bibfnamefont {Mathias}\ \bibnamefont {Pierre}},\ }\bibfield  {title}
  {\enquote {\bibinfo {title} {{Vector SIMP dark matter}},}\ }\href@noop {} {\
  (\bibinfo {year} {2017})},\ \Eprint {http://arxiv.org/abs/1707.01434}
  {arXiv:1707.01434 [hep-ph]} \BibitemShut {NoStop}%
\bibitem [{\citenamefont {Daci}\ \emph {et~al.}(2015)\citenamefont {Daci},
  \citenamefont {De~Bruyn}, \citenamefont {Lowette}, \citenamefont {Tytgat},\
  and\ \citenamefont {Zaldivar}}]{Daci:2015hca}%
  \BibitemOpen
  \bibfield  {author} {\bibinfo {author} {\bibfnamefont {N.}~\bibnamefont
  {Daci}}, \bibinfo {author} {\bibfnamefont {Isabelle}\ \bibnamefont
  {De~Bruyn}}, \bibinfo {author} {\bibfnamefont {S.}~\bibnamefont {Lowette}},
  \bibinfo {author} {\bibfnamefont {M.~H.~G.}\ \bibnamefont {Tytgat}}, \ and\
  \bibinfo {author} {\bibfnamefont {B.}~\bibnamefont {Zaldivar}},\ }\bibfield
  {title} {\enquote {\bibinfo {title} {{Simplified SIMPs and the LHC}},}\
  }\href {\doibase 10.1007/JHEP11(2015)108} {\bibfield  {journal} {\bibinfo
  {journal} {JHEP}\ }\textbf {\bibinfo {volume} {11}},\ \bibinfo {pages} {108}
  (\bibinfo {year} {2015})},\ \Eprint {http://arxiv.org/abs/1503.05505}
  {arXiv:1503.05505 [hep-ph]} \BibitemShut {NoStop}%
\bibitem [{\citenamefont {Albuquerque}\ and\ \citenamefont
  {Baudis}(2003)}]{Albuquerque:2003ei}%
  \BibitemOpen
  \bibfield  {author} {\bibinfo {author} {\bibfnamefont {Ivone F.~M.}\
  \bibnamefont {Albuquerque}}\ and\ \bibinfo {author} {\bibfnamefont {Laura}\
  \bibnamefont {Baudis}},\ }\bibfield  {title} {\enquote {\bibinfo {title}
  {{Direct detection constraints on superheavy dark matter}},}\ }\href
  {\doibase 10.1103/PhysRevLett.90.221301} {\bibfield  {journal} {\bibinfo
  {journal} {Phys. Rev. Lett.}\ }\textbf {\bibinfo {volume} {90}},\ \bibinfo
  {pages} {221301} (\bibinfo {year} {2003})},\ \bibinfo {note} {[Erratum: Phys.
  Rev. Lett.91,229903(2003)]},\ \Eprint {http://arxiv.org/abs/astro-ph/0301188}
  {arXiv:astro-ph/0301188 [astro-ph]} \BibitemShut {NoStop}%
\bibitem [{\citenamefont {Mack}\ \emph {et~al.}(2007)\citenamefont {Mack},
  \citenamefont {Beacom},\ and\ \citenamefont {Bertone}}]{Mack:2007xj}%
  \BibitemOpen
  \bibfield  {author} {\bibinfo {author} {\bibfnamefont {Gregory~D.}\
  \bibnamefont {Mack}}, \bibinfo {author} {\bibfnamefont {John~F.}\
  \bibnamefont {Beacom}}, \ and\ \bibinfo {author} {\bibfnamefont {Gianfranco}\
  \bibnamefont {Bertone}},\ }\bibfield  {title} {\enquote {\bibinfo {title}
  {{Towards Closing the Window on Strongly Interacting Dark Matter:
  Far-Reaching Constraints from Earth's Heat Flow}},}\ }\href {\doibase
  10.1103/PhysRevD.76.043523} {\bibfield  {journal} {\bibinfo  {journal} {Phys.
  Rev.}\ }\textbf {\bibinfo {volume} {D76}},\ \bibinfo {pages} {043523}
  (\bibinfo {year} {2007})},\ \Eprint {http://arxiv.org/abs/0705.4298}
  {arXiv:0705.4298 [astro-ph]} \BibitemShut {NoStop}%
\bibitem [{\citenamefont {Kouvaris}\ and\ \citenamefont
  {Shoemaker}(2014)}]{Kouvaris:2014lpa}%
  \BibitemOpen
  \bibfield  {author} {\bibinfo {author} {\bibfnamefont {Chris}\ \bibnamefont
  {Kouvaris}}\ and\ \bibinfo {author} {\bibfnamefont {Ian~M.}\ \bibnamefont
  {Shoemaker}},\ }\bibfield  {title} {\enquote {\bibinfo {title} {{Daily
  modulation as a smoking gun of dark matter with significant stopping
  rate}},}\ }\href {\doibase 10.1103/PhysRevD.90.095011} {\bibfield  {journal}
  {\bibinfo  {journal} {Phys. Rev.}\ }\textbf {\bibinfo {volume} {D90}},\
  \bibinfo {pages} {095011} (\bibinfo {year} {2014})},\ \Eprint
  {http://arxiv.org/abs/1405.1729} {arXiv:1405.1729 [hep-ph]} \BibitemShut
  {NoStop}%
\bibitem [{\citenamefont {Hochberg}\ \emph {et~al.}(2014)\citenamefont
  {Hochberg}, \citenamefont {Kuflik}, \citenamefont {Volansky},\ and\
  \citenamefont {Wacker}}]{Hochberg:2014dra}%
  \BibitemOpen
  \bibfield  {author} {\bibinfo {author} {\bibfnamefont {Yonit}\ \bibnamefont
  {Hochberg}}, \bibinfo {author} {\bibfnamefont {Eric}\ \bibnamefont {Kuflik}},
  \bibinfo {author} {\bibfnamefont {Tomer}\ \bibnamefont {Volansky}}, \ and\
  \bibinfo {author} {\bibfnamefont {Jay~G.}\ \bibnamefont {Wacker}},\
  }\bibfield  {title} {\enquote {\bibinfo {title} {{Mechanism for Thermal Relic
  Dark Matter of Strongly Interacting Massive Particles}},}\ }\href {\doibase
  10.1103/PhysRevLett.113.171301} {\bibfield  {journal} {\bibinfo  {journal}
  {Phys. Rev. Lett.}\ }\textbf {\bibinfo {volume} {113}},\ \bibinfo {pages}
  {171301} (\bibinfo {year} {2014})},\ \Eprint {http://arxiv.org/abs/1402.5143}
  {arXiv:1402.5143 [hep-ph]} \BibitemShut {NoStop}%
\bibitem [{\citenamefont {Cudell}\ \emph {et~al.}(2014)\citenamefont {Cudell},
  \citenamefont {Khlopov},\ and\ \citenamefont {Wallemacq}}]{Cudell:2014wca}%
  \BibitemOpen
  \bibfield  {author} {\bibinfo {author} {\bibfnamefont {Jean-Rene}\
  \bibnamefont {Cudell}}, \bibinfo {author} {\bibfnamefont {Maxim}\
  \bibnamefont {Khlopov}}, \ and\ \bibinfo {author} {\bibfnamefont {Quentin}\
  \bibnamefont {Wallemacq}},\ }\bibfield  {title} {\enquote {\bibinfo {title}
  {{Effects of dark atom excitations}},}\ }\href {\doibase
  10.1142/S0217732314400069} {\bibfield  {journal} {\bibinfo  {journal} {Mod.
  Phys. Lett.}\ }\textbf {\bibinfo {volume} {A29}},\ \bibinfo {pages} {1440006}
  (\bibinfo {year} {2014})},\ \Eprint {http://arxiv.org/abs/1411.1655}
  {arXiv:1411.1655 [astro-ph.HE]} \BibitemShut {NoStop}%
\bibitem [{\citenamefont {Mahdawi}\ and\ \citenamefont
  {Farrar}(2017)}]{Mahdawi:2017cxz}%
  \BibitemOpen
  \bibfield  {author} {\bibinfo {author} {\bibfnamefont {M.~Shafi}\
  \bibnamefont {Mahdawi}}\ and\ \bibinfo {author} {\bibfnamefont {Glennys~R.}\
  \bibnamefont {Farrar}},\ }\bibfield  {title} {\enquote {\bibinfo {title}
  {{Closing the window on $\sim$GeV Dark Matter with moderate ($\sim$$\mu$b)
  interaction with nucleons}},}\ }\href@noop {} {\  (\bibinfo {year} {2017})},\
  \Eprint {http://arxiv.org/abs/1709.00430} {arXiv:1709.00430 [hep-ph]}
  \BibitemShut {NoStop}%
\bibitem [{\citenamefont {Rich}\ \emph {et~al.}(1987)\citenamefont {Rich},
  \citenamefont {Rocchia},\ and\ \citenamefont {Spiro}}]{RICH1987173}%
  \BibitemOpen
  \bibfield  {author} {\bibinfo {author} {\bibfnamefont {J.}~\bibnamefont
  {Rich}}, \bibinfo {author} {\bibfnamefont {R.}~\bibnamefont {Rocchia}}, \
  and\ \bibinfo {author} {\bibfnamefont {M.}~\bibnamefont {Spiro}},\ }\bibfield
   {title} {\enquote {\bibinfo {title} {A search for strongly interacting dark
  matter},}\ }\href {\doibase http://dx.doi.org/10.1016/0370-2693(87)90788-X}
  {\bibfield  {journal} {\bibinfo  {journal} {Physics Letters B}\ }\textbf
  {\bibinfo {volume} {194}},\ \bibinfo {pages} {173 -- 176} (\bibinfo {year}
  {1987})}\BibitemShut {NoStop}%
\bibitem [{\citenamefont {Starkman}\ \emph {et~al.}(1990)\citenamefont
  {Starkman}, \citenamefont {Gould}, \citenamefont {Esmailzadeh},\ and\
  \citenamefont {Dimopoulos}}]{Starkman:1990nj}%
  \BibitemOpen
  \bibfield  {author} {\bibinfo {author} {\bibfnamefont {Glenn~D.}\
  \bibnamefont {Starkman}}, \bibinfo {author} {\bibfnamefont {Andrew}\
  \bibnamefont {Gould}}, \bibinfo {author} {\bibfnamefont {Rahim}\ \bibnamefont
  {Esmailzadeh}}, \ and\ \bibinfo {author} {\bibfnamefont {Savas}\ \bibnamefont
  {Dimopoulos}},\ }\bibfield  {title} {\enquote {\bibinfo {title} {{Opening the
  Window on Strongly Interacting Dark Matter}},}\ }\href {\doibase
  10.1103/PhysRevD.41.3594} {\bibfield  {journal} {\bibinfo  {journal} {Phys.
  Rev.}\ }\textbf {\bibinfo {volume} {D41}},\ \bibinfo {pages} {3594} (\bibinfo
  {year} {1990})}\BibitemShut {NoStop}%
\bibitem [{\citenamefont {Natarajan}\ \emph {et~al.}(2002)\citenamefont
  {Natarajan}, \citenamefont {Loeb}, \citenamefont {Kneib},\ and\ \citenamefont
  {Smail}}]{Natarajan:2002cw}%
  \BibitemOpen
  \bibfield  {author} {\bibinfo {author} {\bibfnamefont {Priyamvada}\
  \bibnamefont {Natarajan}}, \bibinfo {author} {\bibfnamefont {Abraham}\
  \bibnamefont {Loeb}}, \bibinfo {author} {\bibfnamefont {Jean-Paul}\
  \bibnamefont {Kneib}}, \ and\ \bibinfo {author} {\bibfnamefont {Ian}\
  \bibnamefont {Smail}},\ }\bibfield  {title} {\enquote {\bibinfo {title}
  {{Constraints on the collisional nature of the dark matter from gravitational
  lensing in the cluster a2218}},}\ }\href {\doibase 10.1086/345547} {\bibfield
   {journal} {\bibinfo  {journal} {Astrophys. J.}\ }\textbf {\bibinfo {volume}
  {580}},\ \bibinfo {pages} {L17--L20} (\bibinfo {year} {2002})},\ \Eprint
  {http://arxiv.org/abs/astro-ph/0207045} {arXiv:astro-ph/0207045 [astro-ph]}
  \BibitemShut {NoStop}%
\bibitem [{\citenamefont {Chen}\ \emph {et~al.}(2002)\citenamefont {Chen},
  \citenamefont {Hannestad},\ and\ \citenamefont {Scherrer}}]{Chen:2002yh}%
  \BibitemOpen
  \bibfield  {author} {\bibinfo {author} {\bibfnamefont {Xue-lei}\ \bibnamefont
  {Chen}}, \bibinfo {author} {\bibfnamefont {Steen}\ \bibnamefont {Hannestad}},
  \ and\ \bibinfo {author} {\bibfnamefont {Robert~J.}\ \bibnamefont
  {Scherrer}},\ }\bibfield  {title} {\enquote {\bibinfo {title} {{Cosmic
  microwave background and large scale structure limits on the interaction
  between dark matter and baryons}},}\ }\href {\doibase
  10.1103/PhysRevD.65.123515} {\bibfield  {journal} {\bibinfo  {journal} {Phys.
  Rev.}\ }\textbf {\bibinfo {volume} {D65}},\ \bibinfo {pages} {123515}
  (\bibinfo {year} {2002})},\ \Eprint {http://arxiv.org/abs/astro-ph/0202496}
  {arXiv:astro-ph/0202496 [astro-ph]} \BibitemShut {NoStop}%
\bibitem [{\citenamefont {McDermott}\ \emph {et~al.}(2011)\citenamefont
  {McDermott}, \citenamefont {Yu},\ and\ \citenamefont
  {Zurek}}]{McDermott:2010pa}%
  \BibitemOpen
  \bibfield  {author} {\bibinfo {author} {\bibfnamefont {Samuel~D.}\
  \bibnamefont {McDermott}}, \bibinfo {author} {\bibfnamefont {Hai-Bo}\
  \bibnamefont {Yu}}, \ and\ \bibinfo {author} {\bibfnamefont {Kathryn~M.}\
  \bibnamefont {Zurek}},\ }\bibfield  {title} {\enquote {\bibinfo {title}
  {{Turning off the Lights: How Dark is Dark Matter?}}}\ }\href {\doibase
  10.1103/PhysRevD.83.063509} {\bibfield  {journal} {\bibinfo  {journal} {Phys.
  Rev.}\ }\textbf {\bibinfo {volume} {D83}},\ \bibinfo {pages} {063509}
  (\bibinfo {year} {2011})},\ \Eprint {http://arxiv.org/abs/1011.2907}
  {arXiv:1011.2907 [hep-ph]} \BibitemShut {NoStop}%
\bibitem [{\citenamefont {Foot}\ and\ \citenamefont
  {Vagnozzi}(2015)}]{Foot:2014osa}%
  \BibitemOpen
  \bibfield  {author} {\bibinfo {author} {\bibfnamefont {R.}~\bibnamefont
  {Foot}}\ and\ \bibinfo {author} {\bibfnamefont {S.}~\bibnamefont
  {Vagnozzi}},\ }\bibfield  {title} {\enquote {\bibinfo {title} {{Diurnal
  modulation signal from dissipative hidden sector dark matter}},}\ }\href
  {\doibase 10.1016/j.physletb.2015.06.063} {\bibfield  {journal} {\bibinfo
  {journal} {Phys. Lett.}\ }\textbf {\bibinfo {volume} {B748}},\ \bibinfo
  {pages} {61--66} (\bibinfo {year} {2015})},\ \Eprint
  {http://arxiv.org/abs/1412.0762} {arXiv:1412.0762 [hep-ph]} \BibitemShut
  {NoStop}%
\bibitem [{\citenamefont {Emken}\ \emph {et~al.}(2017)\citenamefont {Emken},
  \citenamefont {Kouvaris},\ and\ \citenamefont {Shoemaker}}]{Emken:2017erx}%
  \BibitemOpen
  \bibfield  {author} {\bibinfo {author} {\bibfnamefont {Timon}\ \bibnamefont
  {Emken}}, \bibinfo {author} {\bibfnamefont {Chris}\ \bibnamefont {Kouvaris}},
  \ and\ \bibinfo {author} {\bibfnamefont {Ian~M.}\ \bibnamefont {Shoemaker}},\
  }\bibfield  {title} {\enquote {\bibinfo {title} {{Terrestrial Effects on Dark
  Matter-Electron Scattering Experiments}},}\ }\href {\doibase
  10.1103/PhysRevD.96.015018} {\bibfield  {journal} {\bibinfo  {journal} {Phys.
  Rev.}\ }\textbf {\bibinfo {volume} {D96}},\ \bibinfo {pages} {015018}
  (\bibinfo {year} {2017})},\ \Eprint {http://arxiv.org/abs/1702.07750}
  {arXiv:1702.07750 [hep-ph]} \BibitemShut {NoStop}%
\bibitem [{\citenamefont {Dziewonski}\ and\ \citenamefont
  {Anderson}(1981)}]{DZIEWONSKI1981297}%
  \BibitemOpen
  \bibfield  {author} {\bibinfo {author} {\bibfnamefont {Adam~M.}\ \bibnamefont
  {Dziewonski}}\ and\ \bibinfo {author} {\bibfnamefont {Don~L.}\ \bibnamefont
  {Anderson}},\ }\bibfield  {title} {\enquote {\bibinfo {title} {Preliminary
  reference earth model},}\ }\href {\doibase
  http://dx.doi.org/10.1016/0031-9201(81)90046-7} {\bibfield  {journal}
  {\bibinfo  {journal} {Physics of the Earth and Planetary Interiors}\ }\textbf
  {\bibinfo {volume} {25}},\ \bibinfo {pages} {297 -- 356} (\bibinfo {year}
  {1981})}\BibitemShut {NoStop}%
\bibitem [{\citenamefont {Emken}\ and\ \citenamefont
  {Kouvaris}(2017)}]{Emken:2017qmp}%
  \BibitemOpen
  \bibfield  {author} {\bibinfo {author} {\bibfnamefont {Timon}\ \bibnamefont
  {Emken}}\ and\ \bibinfo {author} {\bibfnamefont {Chris}\ \bibnamefont
  {Kouvaris}},\ }\bibfield  {title} {\enquote {\bibinfo {title} {{DaMaSCUS: The
  Impact of Underground Scatterings on Direct Detection of Light Dark
  Matter}},}\ }\href@noop {} {\  (\bibinfo {year} {2017})},\ \Eprint
  {http://arxiv.org/abs/1706.02249} {arXiv:1706.02249 [hep-ph]} \BibitemShut
  {NoStop}%
\bibitem [{\citenamefont {{U.S. Government Printing
  Office}}(1976)}]{US_atmosphere}%
  \BibitemOpen
  \bibfield  {author} {\bibinfo {author} {\bibnamefont {{U.S. Government
  Printing Office}}},\ }\href
  {https://ntrs.nasa.gov/archive/nasa/casi.ntrs.nasa.gov/19770009539.pdf}
  {\enquote {\bibinfo {title} {{U.S. Standard Atmosphere}},}\ } (\bibinfo
  {year} {1976})\BibitemShut {NoStop}%
\bibitem [{\citenamefont {Kavanagh}\ \emph {et~al.}(2017)\citenamefont
  {Kavanagh}, \citenamefont {Catena},\ and\ \citenamefont
  {Kouvaris}}]{Kavanagh:2016pyr}%
  \BibitemOpen
  \bibfield  {author} {\bibinfo {author} {\bibfnamefont {Bradley~J.}\
  \bibnamefont {Kavanagh}}, \bibinfo {author} {\bibfnamefont {Riccardo}\
  \bibnamefont {Catena}}, \ and\ \bibinfo {author} {\bibfnamefont {Chris}\
  \bibnamefont {Kouvaris}},\ }\bibfield  {title} {\enquote {\bibinfo {title}
  {{Signatures of Earth-scattering in the direct detection of Dark Matter}},}\
  }\href {\doibase 10.1088/1475-7516/2017/01/012} {\bibfield  {journal}
  {\bibinfo  {journal} {JCAP}\ }\textbf {\bibinfo {volume} {1701}},\ \bibinfo
  {pages} {012} (\bibinfo {year} {2017})},\ \Eprint
  {http://arxiv.org/abs/1611.05453} {arXiv:1611.05453 [hep-ph]} \BibitemShut
  {NoStop}%
\bibitem [{\citenamefont {Erickcek}\ \emph {et~al.}(2007)\citenamefont
  {Erickcek}, \citenamefont {Steinhardt}, \citenamefont {McCammon},\ and\
  \citenamefont {McGuire}}]{Erickcek:2007jv}%
  \BibitemOpen
  \bibfield  {author} {\bibinfo {author} {\bibfnamefont {Adrienne~L.}\
  \bibnamefont {Erickcek}}, \bibinfo {author} {\bibfnamefont {Paul~J.}\
  \bibnamefont {Steinhardt}}, \bibinfo {author} {\bibfnamefont {Dan}\
  \bibnamefont {McCammon}}, \ and\ \bibinfo {author} {\bibfnamefont
  {Patrick~C.}\ \bibnamefont {McGuire}},\ }\bibfield  {title} {\enquote
  {\bibinfo {title} {{Constraints on the Interactions between Dark Matter and
  Baryons from the X-ray Quantum Calorimetry Experiment}},}\ }\href {\doibase
  10.1103/PhysRevD.76.042007} {\bibfield  {journal} {\bibinfo  {journal} {Phys.
  Rev.}\ }\textbf {\bibinfo {volume} {D76}},\ \bibinfo {pages} {042007}
  (\bibinfo {year} {2007})},\ \Eprint {http://arxiv.org/abs/0704.0794}
  {arXiv:0704.0794 [astro-ph]} \BibitemShut {NoStop}%
\bibitem [{\citenamefont {Fichet}(2017)}]{Fichet:2017bng}%
  \BibitemOpen
  \bibfield  {author} {\bibinfo {author} {\bibfnamefont {Sylvain}\ \bibnamefont
  {Fichet}},\ }\bibfield  {title} {\enquote {\bibinfo {title} {{Quantum Forces
  from Dark Matter and Where to Find Them}},}\ }\href@noop {} {\  (\bibinfo
  {year} {2017})},\ \Eprint {http://arxiv.org/abs/1705.10331} {arXiv:1705.10331
  [hep-ph]} \BibitemShut {NoStop}%
\bibitem [{\citenamefont {Baryakhtar}\ \emph {et~al.}(2017)\citenamefont
  {Baryakhtar}, \citenamefont {Bramante}, \citenamefont {Li}, \citenamefont
  {Linden},\ and\ \citenamefont {Raj}}]{Baryakhtar:2017dbj}%
  \BibitemOpen
  \bibfield  {author} {\bibinfo {author} {\bibfnamefont {Masha}\ \bibnamefont
  {Baryakhtar}}, \bibinfo {author} {\bibfnamefont {Joseph}\ \bibnamefont
  {Bramante}}, \bibinfo {author} {\bibfnamefont {Shirley~Weishi}\ \bibnamefont
  {Li}}, \bibinfo {author} {\bibfnamefont {Tim}\ \bibnamefont {Linden}}, \ and\
  \bibinfo {author} {\bibfnamefont {Nirmal}\ \bibnamefont {Raj}},\ }\bibfield
  {title} {\enquote {\bibinfo {title} {{Dark Kinetic Heating of Neutron Stars
  and An Infrared Window On WIMPs, SIMPs, and Pure Higgsinos}},}\ }\href@noop
  {} {\  (\bibinfo {year} {2017})},\ \Eprint {http://arxiv.org/abs/1704.01577}
  {arXiv:1704.01577 [hep-ph]} \BibitemShut {NoStop}%
\bibitem [{\citenamefont {Kouvaris}\ and\ \citenamefont
  {Tinyakov}(2010)}]{Kouvaris:2010vv}%
  \BibitemOpen
  \bibfield  {author} {\bibinfo {author} {\bibfnamefont {Chris}\ \bibnamefont
  {Kouvaris}}\ and\ \bibinfo {author} {\bibfnamefont {Peter}\ \bibnamefont
  {Tinyakov}},\ }\bibfield  {title} {\enquote {\bibinfo {title} {{Can Neutron
  stars constrain Dark Matter?}}}\ }\href {\doibase 10.1103/PhysRevD.82.063531}
  {\bibfield  {journal} {\bibinfo  {journal} {Phys. Rev.}\ }\textbf {\bibinfo
  {volume} {D82}},\ \bibinfo {pages} {063531} (\bibinfo {year} {2010})},\
  \Eprint {http://arxiv.org/abs/1004.0586} {arXiv:1004.0586 [astro-ph.GA]}
  \BibitemShut {NoStop}%
\end{thebibliography}
\end{document}